\documentclass[12pt,a4paper]{article}

\usepackage{epsfig}
\usepackage{amssymb}

\newcommand{\be}{\begin{equation}}
\newcommand{\en}{\end{equation}}
\newcommand{\bea}{\begin{eqnarray}}
\newcommand{\ena}{\end{eqnarray}}

\newcommand{\hbo}{\hbox to 1 true cm {\hfill } }
\newcommand{\tr}{\hbox{tr}}

\def\dslash{\partial\kern-.6em\slash}
\def\kslash{k\kern-.5em\slash}
\def\pslash{p\kern-.4em\slash}
\def\Dslash{D\kern-.6em\slash}
\def\Vslash{V\kern-.7em\slash}
\def\vslash{v\kern-.5em\slash}
\def\rslash{r\kern-.5em\slash}
\def\qslash{q\kern-.5em\slash}

\begin{document}
\vglue 1truecm

\vbox{ UNITU-THEP-11/00  \hfill April 3, 2001
}
\vbox{ RCNP-Th00028  \hfill (revised) }


\vfil
\centerline{\large\bf Gauge invariant $Z(2)$ vortex vacuum textures  }
\centerline{\large\bf and the $SU(2)$ gluon condensate  }

\bigskip
\centerline{ Kurt Langfeld$^a$, Ernst--Michael Ilgenfritz$^b$,
Hugo Reinhardt$^a$ }
\vspace{1 true cm}
\centerline{ $^a$ Institut f\"ur Theoretische Physik, Universit\"at
   T\"ubingen }
\centerline{D--72076 T\"ubingen, Germany}

\vspace{.8 true cm}
\centerline{ and }

\vspace{.8 true cm}
\centerline{ $^b$ Research Center for Nuclear Physics, Osaka University }
\centerline{ Osaka 567-0047, Japan }

\vfil
\begin{abstract}
For $SU(2)$ lattice gauge theory, a new $SO(3)$ cooling procedure is proposed 
which removes the $SU(2)/Z(2)$ coset fields from the lattice configurations 
and reveals a $Z(2)$ vortex vacuum texture different from the $P$--vortex content
obtained in the maximal center gauge.
Cooling can be restricted in a renormalization group invariant way by a parameter 
controlling the remaining $SO(3)$ action density. A gauge invariant $Z(2)$ vortex 
vacuum emerges asymptotically if cooling is not restricted. This ``vortex texture'' 
does not support the string tension $\sigma$ or a finite part of it.  
The $SU(2)$ action density associated with the new $Z(2)$ vortex texture 
has a smooth extrapolation to the continuum limit. We propose an 
interpretation as a mass dimension four condensate related to the gluon
condensate featuring in the operator product expansion.

\end{abstract}

\vfil
\hrule width 5truecm
\vskip .2truecm
\begin{quote}

keywords: {\it center vortex, gauge invariance, confinement,
               gluon condensate, operator product expansion. } 

\end{quote}
\eject

\section{Introduction \hfill }
\label{sec:1} 

Yang--Mills theory has become {\it the } candidate theory of strong interactions in the seventies. 
This confidence is based on asymptotic freedom and the validity of perturbation theory at short 
distances. The long distance/strong coupling regime of this theory, however, defies a sufficient 
understanding until now. More precisely, there are now several competing models well established 
on the lattice to incorporate  the low energy properties of this theory. In particular, the origin 
of such an distinguishing feature of strong interactions like quark confinement is still under 
debate. One of the mechanisms, the center vortex picture will play a prominent r\^ole in this paper. 
Our main concern, however, is the intermediate region where the operator product expansion (OPE) 
is a reliable scheme to describe the onset of non--perturbative physics.
We will point out the existence, in $SU(2)$ lattice gluon dynamics, of an alternative center vortex 
structure which has a good chance to explain the nature of the gluon condensate, the most important 
parameter of the OPE.   
 
\vskip 0.3cm 
Relatively early in the history of QCD, 't~Hooft~\cite{tho74} has pointed out that choosing a 
particular gauge might be useful to identify the agents of confinement. In certain Abelian 
gauges~\cite{tho74}, which allow for a residual $U(1)$ gauge degree of freedom, the Yang--Mills 
ground state appears as a dual superconductor~\cite{tho74,mand} where color--magnetic monopoles of
quantized charge play a r\^ole analogous to Cooper pairs in a superconductor. 
After the search for monopole excitations in the lattice vacua of non--Abelian gauge theory 
(via Abelian projection) has started in the late eighties~\cite{kro87,bornyakov}, 
the demonstration of monopole condensation, necessary to generate confinement through a dual 
Meissner effect, followed in the nineties~\cite{polikarpov,kanazawa,pisa,sch99} 
(for a caveat see~\cite{la2000}). This was demonstrated by evaluating the ``disorder parameter'' 
of confinement. Monopole condensation was observed for different Abelian gauges~\cite{pisa}. 

\vskip 0.3cm 
The idea that vortex free energies might serve as an order parameter for confinement was born 
at the end of the seventies. It dates back to another pioneering work by 't~Hooft~\cite{tho78}
and simultaneous work by Aharonov et al.~\cite{aha78} and was recently confirmed by a numerical 
investigation~\cite{kov00}. Yoneya~\cite{Yon78} and Mack al.~\cite{mack} were the first to 
construct $Z(N)$ topological degrees of freedom from gauge invariant variables and pointed out 
that the so--called center vortices play an important role for the confinement of quarks. 
Basically, these vortices are defined by the property that they contribute a non--trivial center 
element $z_{\alpha}\ne 1$ (among the $N_{\mathrm{color}}$--th roots of unity) to the 
Wilson loop if they are non--trivially linked to the latter. Random fluctuations of the vortices 
provide the area law for the Wilson loop, the signature of confinement~\cite{mack,kov98}.

\vskip 0.3cm 
A revival of the vortex picture arose with the construction of the $P$--vortices on the lattice 
which could be defined after choosing the so--called maximal center gauge (MCG)~\cite{deb98} and 
separating out the center elements from the other lattice degrees of freedom, in this case 
by center projection. For the acceptance of the $P$--vortices as physical objects, it was 
essential that evidence could be presented that they are sensible degrees of freedom in the 
continuum limit~\cite{la98}: the area density of the $P$--vortices as well as their binary 
interactions properly extrapolate to the continuum~\cite{la98}. Moreover, the $P$--vortex 
picture of the Yang--Mills ground state also gives an appealing explanation for the finite 
temperature deconfinement phase transition of Yang--Mills theory~\cite{la99,la00} as the 
breakdown of vortex percolation.

\vskip 0.3cm 
Subsequently, it turned out that center projection even {\it without previous gauge fixing} 
reproduces the $Q\overline{Q}$ potential, also at short distances, a finding which has somewhat 
obscured the relevance of the $P$--vortices for forming the string tension~\cite{fab99}. 
Moreover, in this case it was observed that other properties of the corresponding vortices 
(rather than the purely topological features) strongly depend on the bare lattice coupling 
constant (i.e. the lattice spacing), in a way that made it cumbersome to give a continuum 
interpretation of the apparent vortex degrees of freedom~\cite{fab99}. Thus, center projection 
without appropriate gauge fixing seems not to be the right way to avoid the apparent deficiency 
of the need to find an appropriate gauge, to fight with the corresponding technical Gribov 
problem etc.

\vskip 0.3cm
Above, vortices and the center degrees of freedom were discussed exclusively under 
the aspect of confinement. There is a general opinion that the true confiners are random magnetic 
fluxes which have some transversal extendedness (thick vortices)~\cite{faber_et_al}.
The maximal center projection 
(after MCG fixing) however, ends up with a type of $Z(2)$ gauge field configuration whose vortices 
(thin or $P$--vortices) live on the lattice scale $a$ and are meant to localize the thick vortices. 

\vskip 0.3cm
In the present paper we are going to identify another vortex structure related to the underlying 
$Z(2)$ degrees of freedom which is connected with the non--perturbative dynamics at short distances. 
In a loose sense, the $SO(3)$ cooling method we are proposing serves to separate, among the 
$P$--vortices exhibited by center projection, those $P$--vortices which in fact represent extended 
vortices of small non--Abelian action (supported by the $SO(3)$ part of the gauge field) 
referred to as $SO(3)$ vortices, from the 
real scale $a$ vortices residing in the $Z(2)$ part of the gauge group. Only the first ones are 
in the position to condense and are relevant for confinement. 
Restricting cooling to a finite ratio of $SO(3)$ action density to the string tension $\sigma^2$, 
a residual confining 
force can be defined in a lattice--scale independent way such that the string tension 
is conserved only outside a certain cooling radius and $SO(3)$ vortices 
with a thickness smaller than this radius are 
wiped out. Infinite cooling leaves us with the scale $a$ (thin) $Z(2)$ vortex component alone. 

\vskip 0.3cm
This seems to be a good starting point to establish a connection between the $Z(2)$ gauge field 
content of $SU(2)$ lattice gluon dynamics and the gluon condensate which (at least in the real 
world of $SU(3)$ gauge theory) is an important parameter of particle phenomenology.
A large body of knowledge on low energy properties (resonance physics) and high energy scattering 
is incorporated in the gluon condensate and its short range, non--local structure. In the case of 
the hadronic spectral function approaching intermediate distances
the operator product expansion (OPE)~\cite{shif79} for the current--current correlators was the 
first systematic framework to take into account the non--trivial structure of the vacuum. 
In this approach, non--perturbative properties of the Yang--Mills (or QCD) vacuum are 
parameterized by so--called condensates the values of which are fitted to the experimental 
hadron correlators~\cite{shif79}. 

\vskip 0.3cm
In this paper, we will demonstrate the viability of a new, gauge independent method to separate
low from high energy $SO(3)$ degrees of freedom and to suppress high energy gluons. 
This method exhibits a gauge independent $Z(2)$ center vortex content of the $SU(2)$ vacuum,
different from the 
well--known $P$--vortex structure 
discovered by center projection. 
We propose that the gauge invariant $Z(2)$ vortex component accounts for the average energy 
density after the gluon radiation is subtracted. For this purpose, we will propose cooling 
with respect to the $SO(3)\hat{=} SU(2)/Z(2)$ action. This kind of cooling eliminates the 
$SO(3)$ part of the links which we will refer to in the following as {\it gluons}. 
We find that the $SU(2)$ configurations, asymptotically emerging from the $SO(3)$ cooling procedure 
can be considered as configurations of an effective $Z(2)$ gauge model embedded into $SU(2)$. 
These configurations possess an $SU(2)$ action density which properly extrapolates to the 
continuum limit. We will argue that this action density acquires immediate importance as the mass
dimension four condensate figuring in the OPE as gluon condensate. 

\vskip 0.3cm
The outline of the paper is as follows. 
In the next section we define the new cooling procedure which is based on the $SU(2)/Z(2)$ 
decomposition of the links. Section 3 contains numerical results characterizing the emerging 
gauge invariant $Z(2)$ vortex texture. We show there that these vortices do not contribute 
to the string tension. However, they do account for the gluon condensate as will be shown 
in section 4. There we also briefly review the operator product approach to 
hadronic correlators with a special emphasis on Yang--Mills condensates. The $Z(2)$ vortex 
texture contribution to the gluon condensate is obtained there. We consider the positive 
plaquette model in section 5 focusing on the gluon condensate $O_4$ and find that the latter 
is suppressed by one order of magnitude. Our conclusions are summarized in the final section 6.

\section{The gauge invariant $Z(2)$ vortex vacuum texture \hfill } 
\label{sec:2} 

\subsection{Vortices and gluons \hfill } 
\label{sec:vg} 

Although the problem could be posed also for $SU(3)$ gluon dynamics, in this paper, we will 
concentrate on 4--dimensional pure $SU(2)$ lattice gauge theory.
The partition function
\be 
Z \; = \; \int {\cal D } U \; \exp \biggl\{ - \beta \sum_p s^{W}_p 
\biggr\} \; , 
\label{eq:10} 
\en 
is a high--dimensional integral over $SU(2)$ matrices $U_l=U_{x,\mu}$ associated with the 
links $l=\{x,\mu\}$ of the $L^4$ lattice. The inverse bare lattice coupling constant 
$\beta = 4/g^2$ \footnote{The bare lattice coupling $\beta$ should not be confused with 
the renormalization group $\beta$--function $\bar{\beta}(g)$.} 
is considered to be running with the lattice spacing $a$. Thereby, with $\beta \rightarrow \infty$, 
convergence can be achieved towards the continuum limit for dimensionful physical quantities, 
such as masses, temperatures, string tension, condensates (all given in powers of $a^{-1}$) 
which are extracted from expectation values with respect to the measure (\ref{eq:10}). 
We use the Wilson action with a density given by 
\be 
s^{W}_p \; = \;  1 \, - \, \frac{1}{2}~\tr~U_p \, \,, 
\label{eq:11} 
\en 
while the full action in (\ref{eq:10}) is a sum over all plaquettes $p$.  $U_p$ is the usual 
ordered product $U_p = {\cal P} \prod_{l \in \partial p} U_l$. 

\vskip 0.3cm
For the following considerations it is useful to introduce, besides the link
variable $U_{x,\mu } \in SU(2)$,  the adjoint link variable 
\be
O^{ab}_{x,\mu} \; = \; \frac{1}{2} \tr \biggl\{ U_{x,\mu}~\tau ^a~
U^{\dagger}_{x,\mu}~\tau^b \biggr\} \; = \; O^{ab}[A^b_{\mu}] \; ,
\hbo O^{ab}_{x,\mu} \in SO(3) \; ,
\label{eq:14a}
\en
which do not feel the center degrees of freedom of the links. 
These adjoint links transform under gauge transformations 
according 
\be 
O^{\omega }_{x,\mu} \; = \; \omega_x \; O_{x,\mu} \; \omega ^T_{x+\hat{\mu}} \; ,
\hbo 
\omega ^{ab}_x \; = \; \frac{1}{2}~\tr~\biggl\{ \Omega_x \tau ^a
\Omega^\dagger_x \tau ^b \biggr\} \; .
\label{eq:19}
\en 
Comparing this transformation property with the one of continuum 
gluon fields, i.e. 
\be 
A^{a \, \prime }_{\mu }(x) \; = \; \omega ^{ab}(x) \,  A^b _\mu (x) 
\; + \; \epsilon ^{aef} \, \omega ^{ec}(x) \, \partial _\mu \omega ^{fc}(x) \; , 
\label{eq:19a} 
\en 
we are led to the identification of the gluon fields as the algebra fields 
of the adjoint representation, i.e. 
\be 
O_{x ,\mu }^{ab} \; =: \; \biggl[ \exp \bigl\{ \epsilon ^f 
\, A^f_\mu (x) \, a \bigr\} \, \biggr] ^{ab} \; , 
\hbo \big(\epsilon ^f \bigr)^{ac} := \epsilon ^{afc} \; , 
\label{eq:19b} 
\en 
where $a$ is the lattice spacing. Here we here propose to distinguish 
between the gluon fields which span the SO(3) subgroup and 
the residual $Z_2$ vortex degrees of freedom.

\subsection{Revealing the $Z(2)$--vortex vacuum structure \hfill }
\label{sec:vvs}

In order to detect the inherent effective $Z(2)$ gauge model structure, we will remove 
the gluonic (coset) degrees of freedom from $SU(2)$ configurations by an appropriate
cooling procedure. For this purpose, we define a gluonic action density per link by
\be
s^{gl}_{x,\mu} \; = \; \sum_{\bar{\nu } \not=  \pm \mu } \left\{
1 \; - \; \frac{1}{3} \, \tr _A \,
O_{x,\mu \bar{\nu } } \right\} \; = \; \frac{1}{3}
\sum _{\bar{\nu } \not= \pm \mu } F^a_{\mu \bar{\nu }} [A]
~F^a_{\mu \bar{\nu } }[A]~a^4 \, + \, {\cal O}(a^6) \; ,
\label{eq:30}
\en
where $O_{x,\mu\nu}$ is the plaquette calculated in terms of the $SO(3)$ link elements 
$O_{x,\mu}$ (\ref{eq:14a}).  The sum over $\bar{\nu}$ runs over all forward and backward 
directions orthogonal to $\mu$.  $F^a_{\mu\nu}[A]$ is the (continuum) field strength  
of the (continuum) gluon fields $A_\mu(x)$ and $a$ is the lattice spacing.

\vskip 0.3cm
The new cooling is performed by reducing the gluonic action, i.e. 
by minimization of $s^{gl}_{x,\mu}$ (\ref{eq:30}) with respect to the fields $O_{x,\mu}$. 
Inspired by Ref.~\cite{per00}, we include a kind of self--restriction in our method.
Further cooling of the adjoint link $O_{x\mu}$ is rejected iff the gluonic action is smaller 
than some threshold value
\be
s^{gl}_{x,\mu} \; < \; 8 \kappa ^4 \, a^4 \; .
\label{eq:31}
\en
Thereby $\kappa$ is a gauge invariant cooling scale of mass dimension one. For $\kappa=0$, 
the cooling procedure completely removes the gluon fields from the $SU(2)$ lattice configurations 
leaving only gauge equivalents of $O_{x,\mu}=1$. Notice that the standard cooling method 
(self--restricted or not) minimizes the full $SU(2)$ action, thereby affecting center as well as
coset degrees of freedom. Another crucial difference to the method in~\cite{per00} is that the 
cooling scale there measures the distance of the lattice configurations from classical solutions
(instantons) while in our case the cooling scale constrains the action of the gluon fields.

\vskip 0.3cm
In practice, the cooling procedure works as follows in the $SU(2)$ manifold. The gluonic action 
density $s^{gl}$ (\ref{eq:30}) can be written in terms of the $SU(2)$ fundamental plaquette 
$U_{x,\mu\nu}$, i.e.
\be
s^{gl}_{x,\mu}  \; = \; \frac{4}{3} \sum_{\bar{\nu } \not=  \pm \mu }
\left\{~1 - \left( \frac{1}{2} \, \tr \,
U_{x,\mu \bar{\nu } } \, \right)^2 \; \right\} \; .
\label{eq:32}
\en
A local (maximal) cooling step amounts to a replacement of the
link $U_{x\mu}$ by the cooled variable $U^c_{x,\mu}$
\bea
U^c_{x,\mu} &=& \lambda \, \sum _{\bar{\nu} \not= \pm \mu }
B_{x,\bar{\nu}\mu} \; \left( \frac{1}{2} \, \tr \, U_{x,\mu\bar{\nu}}
\right) \; ,
\label{eq:33} \\
B_{x,\bar{\nu}\mu} &:=& U_{x,\bar{\nu}} \, U_{x+\hat{\bar{\nu}},\mu} \,
U^{\dagger }_{x+\hat{\mu},\bar{\nu}} \,
\label{eq:34}
\ena
where $\lambda$ is a
Lagrange multiplier ensuring
$U^{c \, \dagger}_{x,\mu} \, U^c_{x,\mu} = 1$.
This local cooling step is disregarded iff
\be
1 \; - \; \frac{1}{2} \, \tr \biggl[ U_{x,\mu} \, \left(U^c_{x,\mu} \right)
^\dagger \biggr] \; < \; \kappa ^4  a^4 \; .
\label{eq:35}
\en
The equations (\ref{eq:33}-\ref{eq:35}) define the cooling procedure to be applied in our 
investigations reported below. Taking into account that the normalization $\lambda$ is given by
$\lambda = 1/6 + {\cal O}(a^4) $, the condition (\ref{eq:35}) agrees with (\ref{eq:31}) up to 
order ${\cal O}(a^6)$.  One cooling sweep consists of updating once all links of the lattice
in sequential order according (\ref{eq:33}), i.e. $U_{x,\mu} \rightarrow U^c_{x,\mu} $.
After a finite number of cooling sweeps, the local constraint (\ref{eq:35}) is satisfied all 
over the lattice implying that there is no change in the link variables $U^c_{x,\mu }$ by 
further cooling steps. This is how the cooling procedure stops.

\subsection{Gauge invariance of the texture \hfill }
\label{sec:git}

\begin{figure}[t]
\centerline{
\epsfxsize=10cm
\epsffile{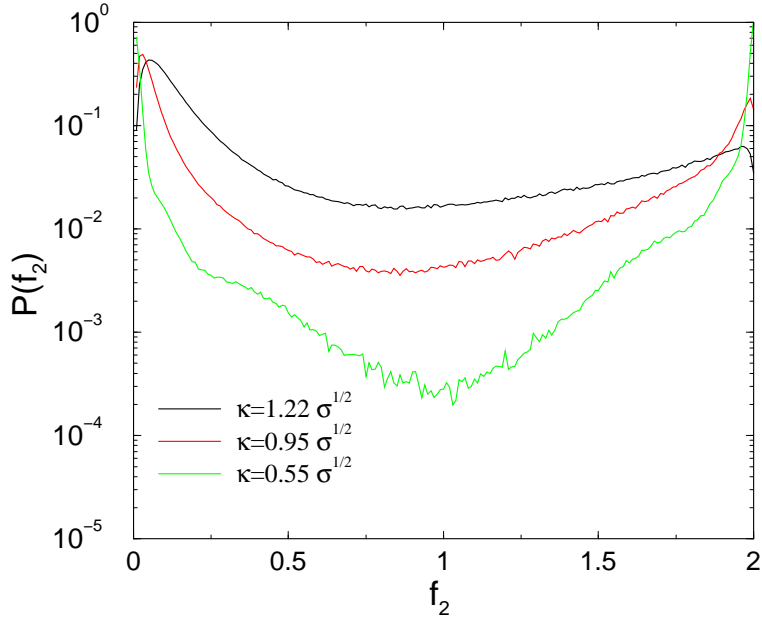}
}
\caption{Separation of the $SU(2)$ action into gluonic radiation (small $f_2$)
               and vortex vacuum texture (large $f_2$).}
\label{fig:2}
\end{figure}
This cooling procedure amounts to a minimization of the $SO(3)$ action density as far as 
tolerated by the parameter $\kappa$, and brings the $SU(2)$ plaquettes as close as possible 
to $\pm 1 $. In the limit $\kappa \rightarrow 0$, this cooling eliminates the $SO(3)$ part 
of the link variables completely. Hence, the remaining field configuration can be viewed as if
generated by an underlying effective $Z(2)$ gauge theory. The field configurations of the latter
are thin center vortices.
Thus, the above cooling procedure 
extracts, in the limit $\kappa \rightarrow 0$, a structure that we call the $Z(2)$ center vortex 
content (``vortex texture'') of a given $SU(2)$ lattice configuration. These vortices are given 
by co--closed manifolds of plaquettes equal to $-1$ and carry large Wilson action.
The standard Wilson action density can be used as a detector for the $Z(2)$ vortex texture, 
even at finite $\kappa$. Although the quantity which will be used for identifying the vortices 
is gauge invariant, one has to make sure that the {\it same} vortex structure is obtained by 
the cooling procedure of subsection \ref{sec:vvs} when two different but gauge equivalent link
configurations $\{ U^{\Omega }_{x,\mu} \}$, $\{ U_{x,\mu} \}$, are analyzed. In order to see this,
one firstly notes that the staples $B_{x,\hat{\nu}\mu }$ (\ref{eq:34}) transform homogeneously, i.e.
\be
U_{x,\mu} \rightarrow U^\Omega_{x,\mu} \; : \hbo
B_{x,\hat{\nu}\mu } \rightarrow B^\Omega_{x,\hat{\nu}\mu}
\; = \; \Omega_x \, B_{x,\hat{\nu}\mu} \, \Omega^\dagger_{x+\hat{\mu}}
\; .
\label{eq:36}
\en
Since the trace of the plaquette is gauge invariant, one finds that the cooled configurations 
obtained from $\{U^\Omega_{x,\mu}\}$ and $\{U_{x,\mu}\}$, respectively, differ by the same gauge 
transformation
\be
U_{x,\mu}  \rightarrow U^c_{x,\mu} \; , \hbo
U^\Omega _{x,\mu}  \rightarrow \Omega_x \, U^c_{x,\mu}
\Omega^\dagger_{x+\hat{\mu}} \; .
\label{eq:37}
\en
The cooling procedure is thus gauge covariant and the distribution of gauge invariant quantities 
calculated on cooled configurations is independent of which gauge copy of the initial field 
configuration cooling is applied to. 

\vskip 0.3cm
In summary, the above introduced $SO(3)$ cooling facilitates a gauge invariant detection of 
the $Z(2)$ center vortex content of (embedded in) a $SU(2)$ lattice configuration. 
As will become clear in the following, this vortex structure does not coincide with the 
$P$--vortices extracted by center projecting links after MCG fixing. If there is coincidence 
with part of the latter, this part becomes insignificant in the continuum limit.
We mentioned in the Introduction, that in the limit $\beta \rightarrow \infty$ the number of 
(dual plaquettes forming) $P$--vortices scales like $a^2$ while we will find here that the 
corresponding number for gauge invariant $Z(2)$ vortices scales like $a^4$
(for fixed $\kappa/\sqrt{\sigma}=O(1)$ and in the limit $\kappa \rightarrow 0$). 
This is because our cooling method also removes thick vortices which have a proper support 
within the $SO(3)$ subgroup. One may speculate that under cooling with respect to the $SO(3)$ 
part of the action, thick vortices disappear by growing in transversal extension. 
Opposite to this, center projection (after MCG) converts thick vortices into thin ones which 
can then be detected by a large $Z(2)$ action density at single plaquettes. This is the reason 
why in the following we will call thick, confining vortices also $SO(3)$ vortices.

\section{Numerical results \hfill }
\label{sec:3}

\subsection{Clustering of $SU(2)$ action \hfill }
\label{sec:ca}

\bigskip
\begin{figure}[t]
\centerline{
\epsfxsize=12cm
\epsffile{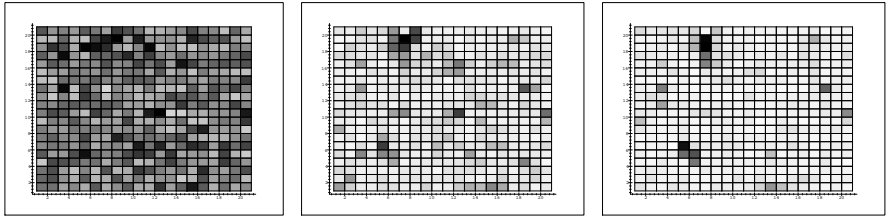}
}
\caption{Distribution of the Wilson action density on 2--dimensional
   hypersurface after cooling steps with $\kappa = \kappa _\mathrm{max}$, 
   $\kappa = 620 \, $MeV, 
   $\kappa = 440 \, $MeV.}
\label{fig:3}
\end{figure}
Generally, the results of $SO(3)$ cooling depend on the cooling scale $\kappa$, given in units 
of the lattice spacing. In order to express the cooling scale in physical units, we will relate 
it to the string tension, given also in lattice units. To get this dimensionless ratio, we will 
adopt the asymptotic scaling law in the 1--loop form
\be
\sigma \, a ^2 (\beta ) \; = \; 0.12 \,
\exp \left\{ - \frac{6 \pi ^2 }{ 11 } \, \left( \beta - 2.3 \right)
\right\} \; ,  \hbo \beta \ge 2.1 \; ,
\label{eq:40}
\en
which fixes the lattice spacing $a$ in units of the string tension $\sigma$ for large enough 
$\beta$. 

In the following, we will employ the trace of the energy momentum tensor 
$\theta ^\mu _{\phantom{\mu } \mu } $ to exhibit the vortex structure of the vacuum. 
In the present case of a $SU(2)$ Yang--Mills theory, this tensor is proportional to the
Wilson action density $s^{W}_p$ (\ref{eq:11}) (see e.g.~(\ref{eq:2.9})).
We denote the Wilson action carried by an arbitrary plaquette by 
$f_2=1 - \frac{1}{2}~\tr~U_{x,\mu\nu}$. Let $P(f_2)$ be the corresponding 
$1$--plaquette probability distribution. This distribution is shown in figure~\ref{fig:2} 
where the data come from a simulation on a $12^4$ lattice at $\beta = 2.3$.
The peak at small values near $f_2 \sim 0$ can be 
attributed to gluon radiation, which would be eliminated by cooling completely only in the 
limit $\kappa/\sqrt{\sigma}\rightarrow 0$. This shows how the constraint (\ref{eq:35}) allows 
to control the action contained in the form of gluon radiation. As expected, it decreases with 
stronger cooling (decreasing energy scale $\kappa$). On the other hand, the peak at $f_2 \sim 2$ 
also grows with increasing cooling. This shows that the contribution of the emerging gauge 
invariant $Z(2)$ vortex texture to the total action density becomes more and more important with 
stronger and stronger cooling. As expected, the $SU(2)$ action density concentrates on single 
plaquettes in the limit $\kappa \rightarrow 0$ which are forming the singular vortex vacuum 
skeleton.  Figure \ref{fig:3} shows the space--time distribution of the $SU(2)$ action density 
on a 2--dimensional hypersurface of a generic configuration generated on a $20^4$ lattice at 
$\beta =2.3$. In this plot, the black spots correspond to the maximum value of the action density 
found on this 2--dimensional hypersurface. One clearly observes how $SO(3)$ cooling leads to the 
clustering of action density at points where the $Z(2)$ vortices pierce the considered hypersurface.

\subsection{Does the gauge invariant $Z(2)$ vortex texture \\ 
                 contribute to the string tension ?\hfill }
\label{sec:vdst}

\begin{figure}[t]
\centerline{
\epsfxsize=12cm
\epsffile{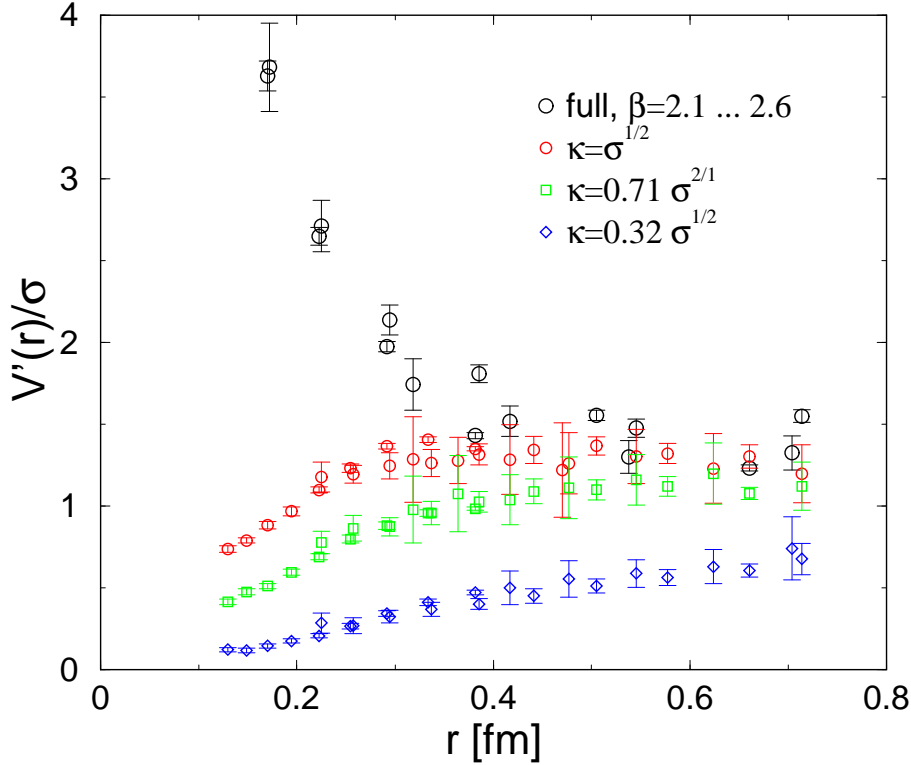}
}
\caption{ The static quark $Q\overline{Q}$ force as function of the 
         distance $r$ between 
         quark and anti--quark for full non--Abelian and $SO(3)$ 
         cooled configurations,
         at different $\kappa/\sqrt{\sigma}$ 
         checked with respect to scaling for various $\beta$. }
\label{fig:4}
\end{figure}
Let us now investigate the relation between the center vortices 
of the MCG projection ($P$--vortices)~\cite{deb98} and the gauge invariant $Z(2)$ vortex texture 
which is defined by the $Z(2)$ gauge fields remaining after the $SO(3)$ cooling described above. 
The contribution of these $Z(2)$ vortices to the string tension is useful to look at. It was 
established that configurations the links of which were projected onto center elements after 
MCG fixing -- the $P$--vortex configurations -- reproduce the full string tension $\sigma $ to 
good accuracy (for most recent results, which also cover a discussion of the practical Gribov 
problem see~\cite{bor00,ber00,sta00}). 

\vskip 0.3cm 
For comparison, we investigate the static $Q\overline{Q}$ attractive force and contrast the 
result from cooled configurations (at different cooling scales representing the remaining 
$SO(3)$ action) to the result obtained in full $SU(2)$ Yang--Mills theory. In each case, we 
used $\beta$--values ranging from $2.1$ to $2.6$ to check for scaling of the force as a function 
of the distance. Our results for a $12^4$ lattice are shown in figure \ref{fig:4}.
We find that the data points (for no cooling and a given cooling scale) 
obtained at different $\beta$ fall on top of the same curve, respectively, thus establishing 
proper scaling also for the potential gained from cooled configurations. 
This shows the advantage to have a renormalization group invariant formulation for restricted 
cooling. 
We find that the cooling procedure strongly affects the force in a range of distances growing 
with decreasing cooling scale $\kappa/\sqrt{\sigma}$. At short distance this was expected since 
the behavior at small $r$ is dominated by the exchange of gluons, which is already partially (cf.  
figure \ref{fig:2}) eliminated by cooling. Moreover, for stronger cooling (smaller cooling scale
$\kappa $) 
the value of the full string tension is approached only asymptotically, such that lowering the 
cooling scale $\kappa/\sqrt{\sigma}$ shifts the asymptotic region to larger distances $r$.
The explanation is that $SO(3)$ cooling washes out the $SO(3)$ vortices. 
We conclude that the gauge invariant $Z(2)$ vortex texture (which survives in the limit 
$\kappa \rightarrow 0$ of unrestricted cooling) is not related to the confinement property of 
the $SU(2)$ vacuum. In particular, it is not identical with the $P$--vortex ensembles constructed 
by means of center projection.

\vskip 0.3cm 
After the major part of the confining vortices 
is removed by cooling, it remains to be clarified which specific physical significance 
could be assigned to the exposed $Z(2)$ vortex ensembles. 
Following the original version of this paper, it was observed 
in~\cite{la00b} that the masses of the low-lying glueballs, $O^+$ and $2^+$, are 
rather insensitive against the here proposed cooling. Below, we will give 
further arguments supporting 
the idea that these configurations are not just lattice artefacts but play an important r\^ole 
in the non--perturbative physics at intermediate distances dealt with in the OPE.

\section{Hadronic correlation functions and OPE\hfill }
\label{sec:4}

\subsection{The operator product expansion }
\label{sec:ope}

\begin{figure}[t]
\centerline{
\epsfxsize=12cm
\epsffile{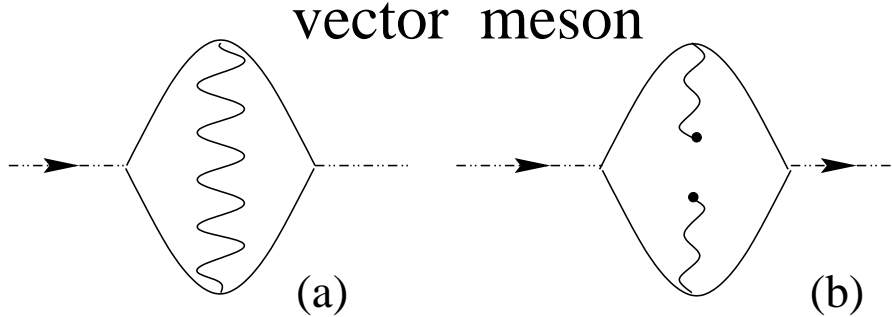}
}
\caption{Two Feynman diagrams contributing to the current--current correlation function in the 
         vector meson channel; radiative gluon (a) and condensate contribution (b). }
\label{fig:1}
\end{figure}
In this section, we will briefly review the operator product expansion (OPE),
which is the framework to discuss the properties of hadronic 
resonances~\cite{shif79}, and the inherent ambiguities. To be specific, 
consider the hadronic current correlation function in the vector meson channel, 
\be
M^{\mu \nu }(q) \; = \; i \int d^4x \; e^{iqx} \; \langle
\Omega \vert \, T \, j^{\mu }(x) \, j^\nu (0) \, \vert \Omega \rangle \; ,
\label{eq:2.1}
\en
where $\vert \Omega \rangle $ denotes the true ground state of Yang--Mills
theory. Current conservation implies
\be
M^{\mu \nu }(q) \; = \; \left(q^\mu q^\nu \; - \; q^2 \, g^{\mu \nu }
\right) \; M(q^2) \; .
\label{eq:2.2}
\en
According to Wilson~\cite{wil69}, the bilocal, time--ordered product of
two operators at short distance can be written in terms of local operators
\be
T \, j^{\mu }(x) \, j^\nu (0)
\; = \; \sum _{n=0}^\infty C_n(x) \; \hat{O}_n(0)  \; , \hbo \hat{O}_0 = 1 \; ,
\label{eq:2.3}
\en
where $n$ labels the canonical mass dimension of the local operators 
$\hat{O}_n$. The Wilson coefficients $C_n(x)$ contain the singular behavior
occurring when the point splitting is removed as $x \rightarrow 0$.
The vacuum expectation values of the operators $\hat{O}_n$ correspond to 
{\it physical} observables, which are called condensates.
Hence, sandwiching (\ref{eq:2.3}) with an arbitrary trial state
$\vert \psi \rangle $ and resorting to a dimensional analysis, one finds
\be
M(q^2) \; = \;  C^\psi _0(q^2) \, 1^\psi \;
+ \; \frac{ O^\psi _2 }{ q^2 }
\; + \; \frac{ O^\psi _4 }{ q^4 } \; + \; {\cal O}\left( 1/q^6 \right) \; ,
\hbo O^\psi _n := \langle \psi \vert \hat{O}_n \vert \psi \rangle \; .
\label{eq:2.4}
\en
Since an infinite number of degrees of freedom is encoded in the
wave function $\vert \psi \rangle $, the matrix element of the unit
operator, $1^\psi = \langle \psi \vert \hat{1} \vert \psi \rangle$
is generically depending on the state under discussion.
Usually this is included in the coefficient function. Therefore the function
$C_0^\psi(q^2)$ in (\ref{eq:2.4}) does depend on the state
(see below and~\cite{la95} for an illustration). 
This fact is just a reflection of the scale anomaly~\cite{col77}. 
The other Wilson coefficient functions are defined in (\ref{eq:2.3}) 
without reference to a particular state.
Note also that a logarithmic dependence of $C^\psi _0(q^2)$
on $q^2$ is not excluded by the dimensional analysis. In particular,
choosing $\vert \psi \rangle = \vert 0 \rangle $, 
the perturbative vacuum satisfying $O_n^0 =0$ for $n>0$, yields
\be
C^0_0(q^2) \; = \; M_{\rm per} (q^2)  \;  ,
\label{eq:2.6}
\en
where $M_{pert}$ is the correlator (\ref{eq:2.2}) calculated by summing
perturbative diagrams up to a finite order using tree--level propagators and
vertices. In this case, one finds up to second order in the gauge coupling 
constant $g$
\be
C^0_0(q^2) \; = \;  - \frac{1}{8 \pi ^2 } \left( 1 + \frac{\alpha _s }{ \pi }
\right) \; \ln \frac{q^2}{\nu ^2 } \; + \; {\cal O}\left( \alpha _s ^2
\right) \; ,
\label{eq:2.5}
\en
where $\nu $ is the renormalization point and $\alpha_s = g^2 /4\pi$.

\vskip 0.3cm
It has been known for some time that the condensates are ambiguous since
a splitting like (\ref{eq:2.4}) into a so--called perturbative part and
contributions from the OPE corrections is not well--defined (see 
e.g.~\cite{zak92}). This conclusion is based on the observation that a 
partial re--summation of perturbative diagrams to all orders
can generate, after renormalization, contributions $\exp(-\frac{1}{g^2})$ 
to correlators. Containing essential
singularities in the coupling constant, these contributions are imitating 
OPE condensate corrections (see e.g.~\cite{zak92}). These results do not 
invalidate the considerations above since the re--summation of perturbative 
diagrams to all orders bears the potential to change the properties
of the reference state $\vert \psi \rangle $ under discussion.
This has been illustrated for the two--dimensional Gross--Neveu model
in Ref.~\cite{la95}. The model is solvable in the large $N_f$ limit 
where $N_f$ is the number of fermion flavors. In leading order of the 
$1/N_f$--expansion, the ladder re--summation of perturbative diagrams 
results in an exact, non--perturbative gap equation which admits a non--trivial 
solution corresponding to the {\it true} ground state for 
$N_f \rightarrow \infty$, which we call $\vert BCS \rangle $.
Doing this re--summation of all perturbative diagrams which contribute
in the large $N_f$ limit yields 
\be
M(q^2) \; = \;  C^{BCS}_0(q^2) \; + \; \frac{ O^{BCS}_2 }{ q^2 }
\; + \; \frac{ O^{BCS} _4 }{ q^4 } \; + \; {\cal O}\left( 1/q^6 \right) \; .
\label{eq:2.7}
\en
In this example, the numbers $O^{BCS}_n$, which vanish in the
infinite $N_f$ limit, account for correlations which are {\it not
present} in the wave function $\vert BCS \rangle $ and which become 
non--negligible at finite $N_f$. Moreover, $C^{BCS}_0(q^2) $
contains already the power--like OPE corrections.
If one does not perform the summation over an infinite number of diagrams, 
one obtains from the OPE
\be
M(q^2) \; = \;  C^0 _0(q^2) \; + \; \frac{ O^0 _2 }{ q^2 }
\; + \; \frac{ O^0 _4 }{ q^4 } \; + \; {\cal O}\left( 1/q^6 \right) \; ,
\label{eq:2.6a}
\en
where $C^0 _0(q^2)$ is calculated from a large but {\it finite }
number of perturbative diagrams (see (\ref{eq:2.6})).
The superscript $0$ signals that the condensates $O_n^0 $
carry all the information on the true vacuum state, since
contributions which could imitate a condensate are lacking in $C^0 _0(q^2)$.
In particular, the parameter $O^0_2$ reflects the dynamical
generated fermion mass which is present in the true vacuum of the infinite
$N_f$ limit but goes beyond perturbation theory. 

\vskip 0.3cm
Both (\ref{eq:2.6a}) and (\ref{eq:2.7}), respectively, represent the 
{\it unique} answer for $M(q^2)$ at large values of $q^2$. The example of 
the Gross--Neveu model demonstrates that choosing for $\vert \psi \rangle$
the true ground state $\vert \Omega \rangle $ does not necessarily imply a 
unique definition of the condensates $O^\Omega _n$. In addition,
usually one demands that, with $M_{\rm per}$ given in (\ref{eq:2.5}),
\be
C^\Omega_0(q^2) \; = \; M_{\rm per}(q^2) \; .
\label{eq:2.5a}
\en
which is indeed satisfied in the case of (\ref{eq:2.6a}).
The summation of a finite (even large) number of perturbative
diagrams cannot imitate the renormalization group invariant
dependence of the condensates on the coupling $g$. Therefore, the
choice (\ref{eq:2.5a}) is specific in as far it shifts the maximum content 
of information on the non--trivial properties of the true vacuum 
$\vert \Omega \rangle $ to the condensates $O_n^\Omega $.

\subsection{The Yang--Mills condensate on the lattice}

\label{sec:ymc}

The local condensates $O_n$ entering the OPE (\ref{eq:2.3}) are required to be
physical observables. Hence, in the case of Yang--Mills theory, they consist 
of gauge invariant and renormalization group invariant quantities. In the 
present paper, we will concentrate onto the condensate $O_4$ of 
canonical mass dimension four. 

\vskip 0.3cm 
Using the trace of the energy momentum tensor $\theta ^\mu _{\phantom{\mu } 
\mu } $ for the operator $\hat{O}_4$ meets the gauge invariance condition. 
However, it has been known for a long time that the contribution from 
gluon radiation yields divergent results for the expectation value 
$\langle \theta ^\mu _{\phantom{\mu } \mu } \rangle $, hence, violating 
the second criterion of renormalization group invariance. For later 
convenience, we illustrate this observation for the case of lattice 
regularization. Calculating $\theta  ^\mu _{\phantom{\mu } \mu }$ by studying 
scale variations, one finds
\be
\langle \theta  ^\mu _{\phantom{\mu } \mu } \rangle \; = \;
\frac{1}{a^4} \; \frac{d \beta (a) }{ d \ln a } \;
\langle  1 \, - \, \frac{1}{2}~\tr~U_p \rangle \; ,
\label{eq:2.9}
\en
where $\beta(a)$ is the running bare lattice coupling. 
Using (lattice) perturbation theory one finds that 
$$ 
\langle 1 \, - \, \frac{1}{2}~\tr~U_p \rangle \; = \; 
\frac{3}{4 \beta } \; + \; {\cal O }(1 / \beta^2)  \; . 
$$ 
Given (\ref{eq:40}), this implies that $\langle \theta  ^\mu 
_{\phantom{\mu } \mu } \rangle$ strongly diverges in the continuum limit 
$a \rightarrow 0$. 

\vskip 0.3cm 
In order to arrive at a sensible (i.e.~gauge and RG invariant) definition 
of $O_4$, one closely follows the perturbative approach augmented 
by OPE corrections~\cite{shif79} and defines 
\be 
O_4^{\mathrm{sub}} \; := \; \langle : \theta ^\mu _{\phantom{\mu } \mu } : 
\rangle \; := \; 
\langle \theta ^\mu _{\phantom{\mu } \mu } \rangle _{\rm sub} 
\label{eq:i2} 
\en 
where $\langle \theta ^\mu _{\phantom{\mu } \mu } \rangle _{\rm sub}$ 
denotes the expectation value from which the contribution from gluon 
radiation is subtracted. A technique subtracting perturbative contributions 
was firstly employed in lattice calculations in the early 
eighties~\cite{kri81,leipzig,ilg82}, and was pursued to high order in a 
recent study of the OPE corrections~\cite{bur98}. 

\vskip 0.3cm 
By subtracting perturbative gluonic parts, both, 
the continuum approach which distinguishes between perturbative 
gluon radiation and non--perturbative vacuum properties 
as well as the lattice approach, 
have the shortcoming that they do not specify the remnants of the vacuum 
which form the gluon condensate once gluonic contributions have been 
subtracted {\it to all orders}. 

\vskip 0.3cm 
Here, we define the condensate  
through the subtraction of the contribution of the gluons (i.e. 
subtracting the $SO(3)$ coset fields) by means of the above cooling 
procedure rather than by subtracting perturbation theory, i.e.,
\bea 
O_4 \; & := & \; \lim _{\kappa \rightarrow 0} O_4(\kappa) \; , \\ \nonumber 
O_4(\kappa) a^4 \; & = & \;\frac{ 24 }{ \pi ^2} \, \biggl\langle 1 \; - \;
\frac{1}{2} \; \tr \; 
  U_p \biggr\rangle _{SO(3)~\mathrm{cooled~with~scale}~\kappa} \; . 
\label{eq:41}
\ena
This definition amounts to the choice - in the sense of eq. (\ref{eq:2.5a}) -
\be
C^\Omega_0(q^2) \; = \; M_{\rm SO(3)}(q^2) \; .
\en
We point out that $M_{\rm SO(3)}$ not necessarily contains only 
{\it perturbative} gluon contributions. Hence, in general, 
$M_{\rm SO(3)} \ne M_{\rm per}$. 

\vskip 0.3cm 
This identification offers the appealing feature that the 
condensate of mass dimension four, $O_4$, could be given by the vacuum  
energy density of the $Z_2$ gauge system which remains after suppressing
the contribution of the coset fields by the $SO(3)$ cooling method. 
It remains to show that the condensate $O_4$ (\ref{eq:41}) 
properly extrapolates to the continuum limit $a \rightarrow 0$, 
which will be done in the next subsection. 

\subsection{Numerical results for the $Z(2)$ vortex induced \\
            gluon condensate}
\label{sec:vigc}

\begin{figure}[t]
\centerline{
\epsfxsize=12cm
\epsffile{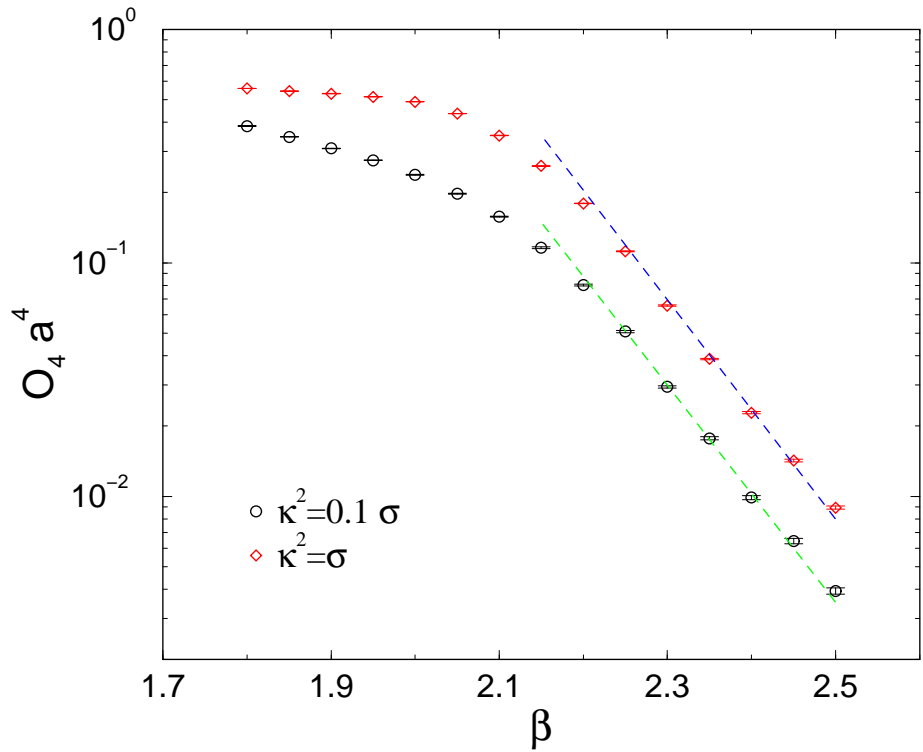}
}
\caption{ $O_4 $ in units of the
   lattice spacing $a$ extracted from $SO(3)$ cooled configuration as
   function of $\beta $. }
\label{fig:5}
\end{figure}
In section~\ref{sec:3}, we have demonstrated a string like clustering of
$SU(2)$ action density which corresponds to a gauge
invariant $Z(2)$ vortex vacuum texture. In this subsection, we will show that,
in addition to gauge invariance, this action density, which is carried
exclusively by the $Z(2)$ gauge fields, possesses the correct renormalization 
group dependence on the $SU(2)$ inverse coupling $\beta$ inherited from the 
full $SU(2)$ lattice gluon dynamics. Hence, this action density can be considered
as a physical observable and can be used to define the mass dimension four condensate 
$O_4$ as outlined in the previous subsection. 

\vskip 0.3cm
Recalling (\ref{eq:41}), $O_4$ should be compared
with what is defined in the literature as the gluon condensate 
\be
\biggl\langle \frac{1}{4 \pi ^2 } :F^a_{\mu \nu } F^a_{\mu \nu } : 
\biggl\rangle \; , 
\label{eq:i3} 
\en
which is obtained from fits to the hadronic spectral function~\cite{shif79}.
We emphasize, however, that there is abuse of notation in (\ref{eq:i3}) 
since the field strength $F^{a}_{\mu \nu }$ is defined in terms of the 
coset (gluon) fields which do not contribute to the above defined 
condensate $O_4$ by construction.
First results which employ the Wilson loop for extracting the 
gluon condensate were obtained in the early eighties from lattice gauge 
theories~\cite{kri81,leipzig,ilg82}. The most recent value for the 
non--natural case of the pure $SU(2)$ gauge theory is given 
by~\cite{ilg00} 
\be
O_4 \; \approx \; 0.15 \; \mathrm{GeV}^4
\label{eq:41a}
\en
and was calculated from field strength correlation functions 
(for a recent review of the method see e.g.~\cite{lisabon}). 

\begin{figure}[t]
\centerline{
\epsfxsize=12cm
\epsffile{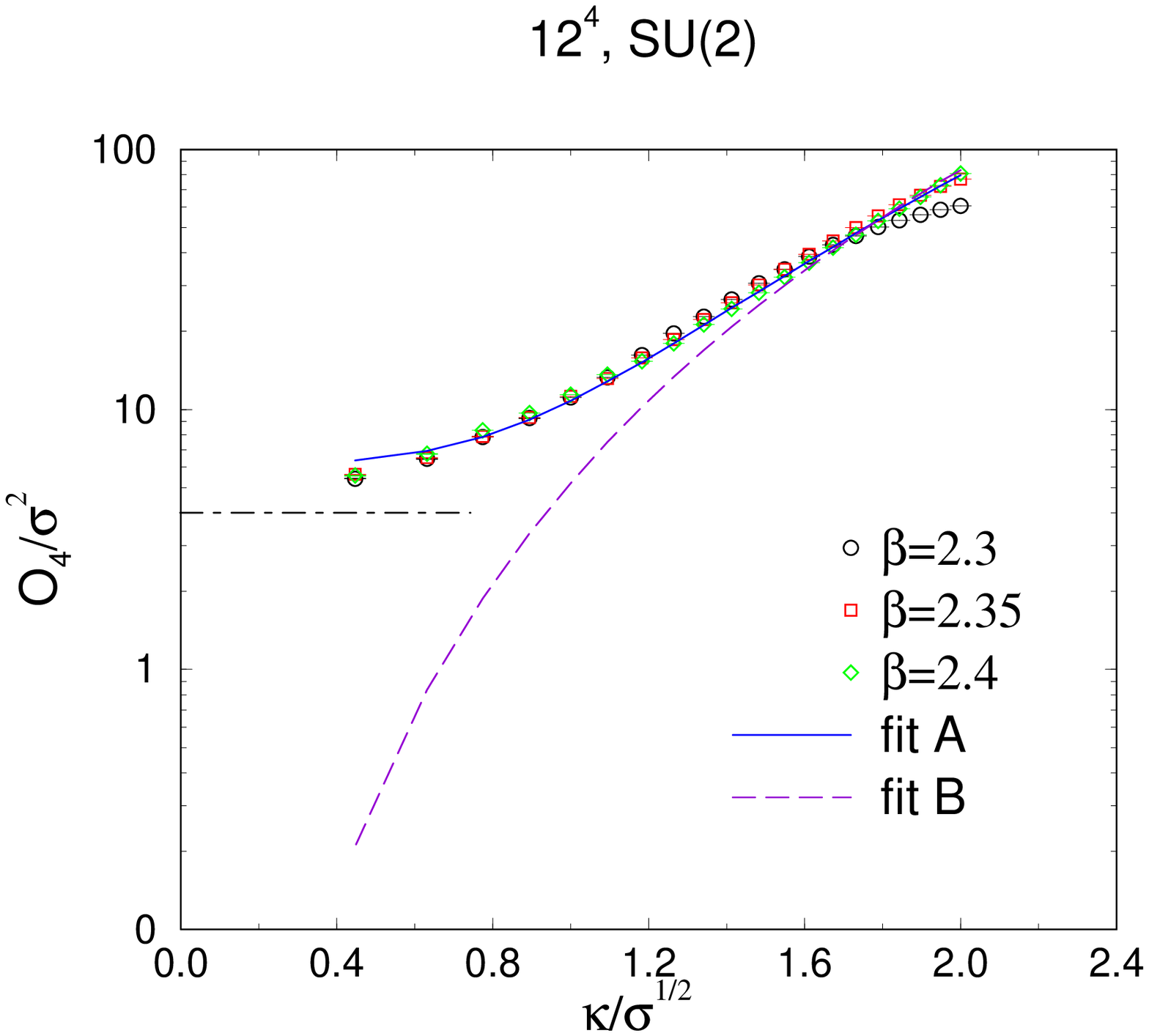}
}
\caption{ The gluon condensate $O_4$ in units of the string tension 
   $\sigma $ squared,
   as function of the cooling scale $\kappa/\sqrt{\sigma}$ (see text). }
\label{fig:6}
\end{figure}

\vskip 0.3cm
In order to assign a physical meaning to the quantity (\ref{eq:41}), 
it is important to check whether this yields a finite and non--vanishing 
value in the continuum limit $a \rightarrow 0$.
We have calculated $f_2 \, a^4= \pi^2 G_2 a^4 /24$ as function of $\beta $
for cooled configurations,  
fixing the cooling scale $\kappa ^2 = 0.5~\sigma $.
The result is shown in figure \ref{fig:5}.
Our results nicely meets with the expectation from perturbative scaling
(dashed line) for $\beta > 2.2$. This observation only tells us that
we can safely extract a signal for the desired expectation value which
remains valid in the continuum limit. 
Again, it turns out advantageous to have the self--restriction of 
the cooling in a scale independent way.

\vskip 0.3cm
In order to show the stripping off of the 
gluonic contributions
from the OPE parameter $O_4$ (\ref{eq:41}), we can follow the limit 
$\kappa \rightarrow 0$. 
The result for $O_4(\kappa)$ (\ref{eq:41}) as function of $\kappa $ given in 
physical units is presented in figure~\ref{fig:6}. Data from various 
$\beta$ in the scaling region $\beta > 2.2$ fall on the same curve. 
For $\kappa > \sqrt{\sigma}$, a significant contribution of the gluon 
radiation to the ''gluon condensate'' is still present, 
and one asymptotically expects $O_4(\kappa) \propto \kappa ^4$. 
This behavior is confirmed by our lattice data (see figure~\ref{fig:6}). 
Roughly at the confining scale, set by $\kappa \approx \sigma ^{1/2}$, 
the contribution of the gluons to $O_4(\kappa)$ becomes comparable
with the gluon condensate carried by the $Z(2)$ vortex vacuum texture. 
For clearly displaying the contribution of the texture, we compare the 
lattice data with the model fits
\bea 
O_4 /\sigma ^2 &=& a_0 \, + \, a_1 \, \kappa^4 \; , \hbox to 3cm {\hfill (fit A) }
\nonumber \\ 
O_4 /\sigma ^2 &=&  \, b_1 \, \kappa^4 \; , \hbox to 3cm {\hfill (fit B)} \; . 
\nonumber 
\ena
In both cases, the terms proportional to $\kappa ^4$ parameterize the 
gluonic contribution while the $a_0$ term of fit A specifies the 
$Z_2$ vortex content. The lattice data clearly favorites fit A. 
We find it convincing that the ``gluon condensate'' $O_4(\kappa)$ 
approaches a finite value in the limit $\kappa \rightarrow 0$ which is 
roughly consistent with the known value (\ref{eq:41a}) (dash--dotted line 
in figure~\ref{fig:6}).

\section{The positive plaquette model (PPM) \hfill }

\begin{figure}[t]
\centerline{
\epsfxsize=12cm
\epsffile{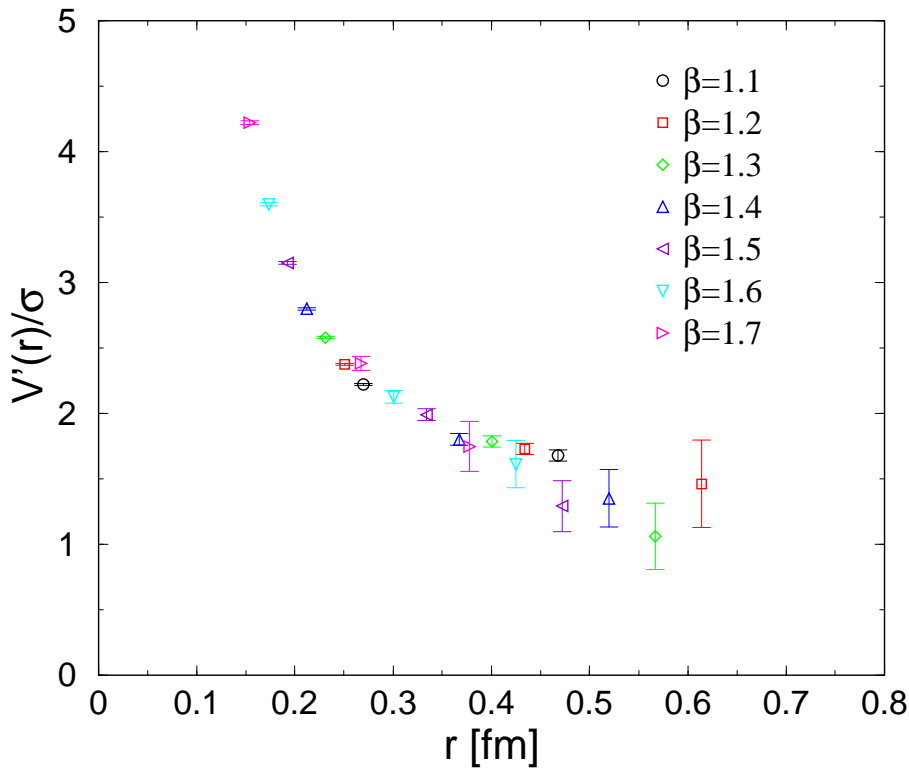}
}
\caption{ The static quark anti--quark force calculated in the PPM. } 
\label{fig:7}
\end{figure}
The numerical results presented above were obtained with the $SU(2)$ Wilson 
action, which includes a definite prescription of the interaction between 
center and coset fields. The question arises how the residual action 
density which is carried by the $Z(2)$ gauge fields after $SO(3)$ cooling 
does depend on the choice of the lattice action used in the simulation. 
In order to get some information on this dependence, we adopt an extreme 
point of view in this section and repeat the 
analysis of the previous section using the positive plaquette model 
(PPM). This model is defined by the partition function
\be 
Z_{\rm ppm}  \; = \; \int {\cal D } U \; \prod_{x \, \mu \nu } 
\; \theta \bigl( \tr~U_{x,\mu\nu} \bigr) \; 
\exp \biggl\{ \frac{\beta}{2} \, \tr~U_{x,\mu\nu} \biggr\} \; . 
\label{eq:p1} 
\en 
The Gibbs weight is that of the Wilson action up to the fact 
that link configurations which would lead to negative plaquettes 
are rejected. One expects that the latter constraint strongly 
influences the asymptotic $Z(2)$ gauge field configurations 
remaining after cooling.  

\vskip 0.3cm 
A set of low energy quantities -- the string tension, the glueball masses
and the topological susceptibility --  
have been studied in great detail within the PPM in~\cite{fin95}. 
It was found that such observables in units of the string tension
become independent 
of the lattice regulator $a$ when the continuum limit is approached
within a certain scaling window $\beta \in [1.3, \, 2.1]$. Moreover, 
the values of the above observables quantitatively agree with the values 
obtained with Wilson action. However, the renormalization group scaling 
of the lattice spacing with $\beta$ does not match with the 
expectations from continuum perturbation theory in the investigated 
scaling window of the PPM. In fact, a scaling 
\be 
\sqrt{\sigma } \, a(\beta ) \; = \; 0.36 \; - \; 0.3 \, \bigl( 
\beta \, - \, 1.3\, \bigr) \;, \hbo 
\beta \in [1.3 , \, 2.1] 
\label{eq:p2} 
\en 
is consistent with the numerical data presented in~\cite{fin95}. 
In order to test our algorithm for simulations of the PPM, 
we have re--calculated the quark--antiquark force in physical units using 
$a(\beta)$ (\ref{eq:p2}) as input. 
We could verify that the data points 
obtained for several values of the lattice spacing fall on top of 
the same curve (see fig.~\ref{fig:7}). 

\vskip 0.3cm
The continuum limit, however, is problematic to assess. 
Notice that we cannot assume that (\ref{eq:p2}) is valid upto $a=0$, 
otherwise the equation suggests a strong coupling UV fix point at 
\be 
g_{\rm fix} \;:= \; g(a \rightarrow 0) \; \approx \; 1.26 \; , 
\hbo (\beta = 4/g^2)
\label{eq:p3} 
\en
and a renormalization group $\beta$--function of 
\be 
\bar{\beta}(g) \; \approx \; \frac{g}{\Lambda} \frac{d\Lambda }{dg} 
\; = \; 1.25 \, - \, \frac{2}{g^2} \; , 
\label{eq:p4} 
\en 
where we have defined the UV cutoff by $\Lambda = \pi/a $. 
Such a fix point contradicts asymptotic freedom. 
The equations (\ref{eq:p3},\ref{eq:p4}) are based on extrapolation of 
numerical data to the continuum limit. Therefore, it cannot be excluded that 
the onset of perturbative physics is postponed to the region $\beta >2.1$. 
Note, however, that at $\beta =2.1$ the UV cutoff is of order 
$\Lambda \approx 11 \, $GeV (if we take the reference scale 
$\sqrt{\sigma } = 440 \, $MeV), where perturbative scaling should dominate. 

\begin{figure}[t]
\centerline{
\epsfxsize=8cm
\epsffile{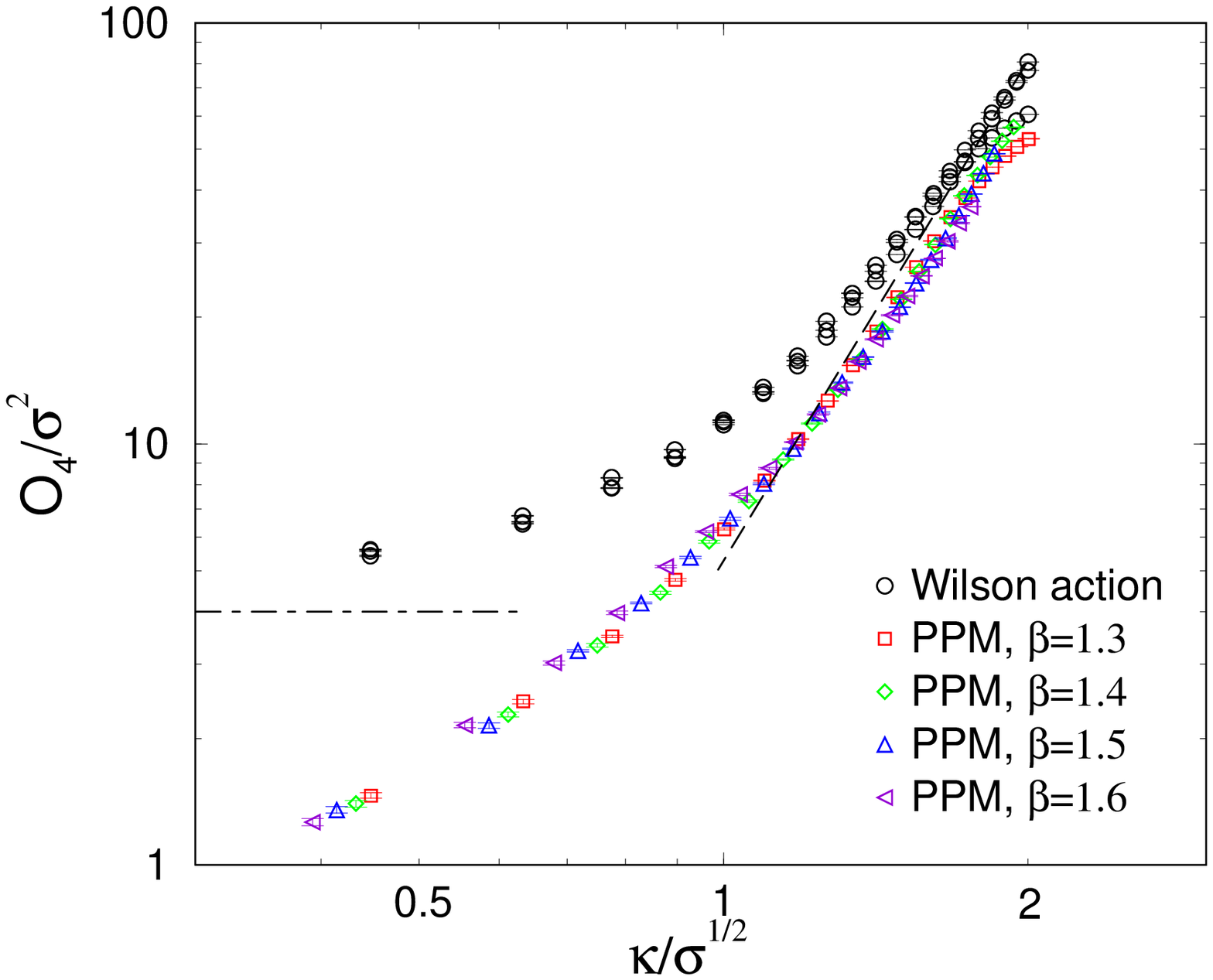}
\epsfxsize=8cm
\epsffile{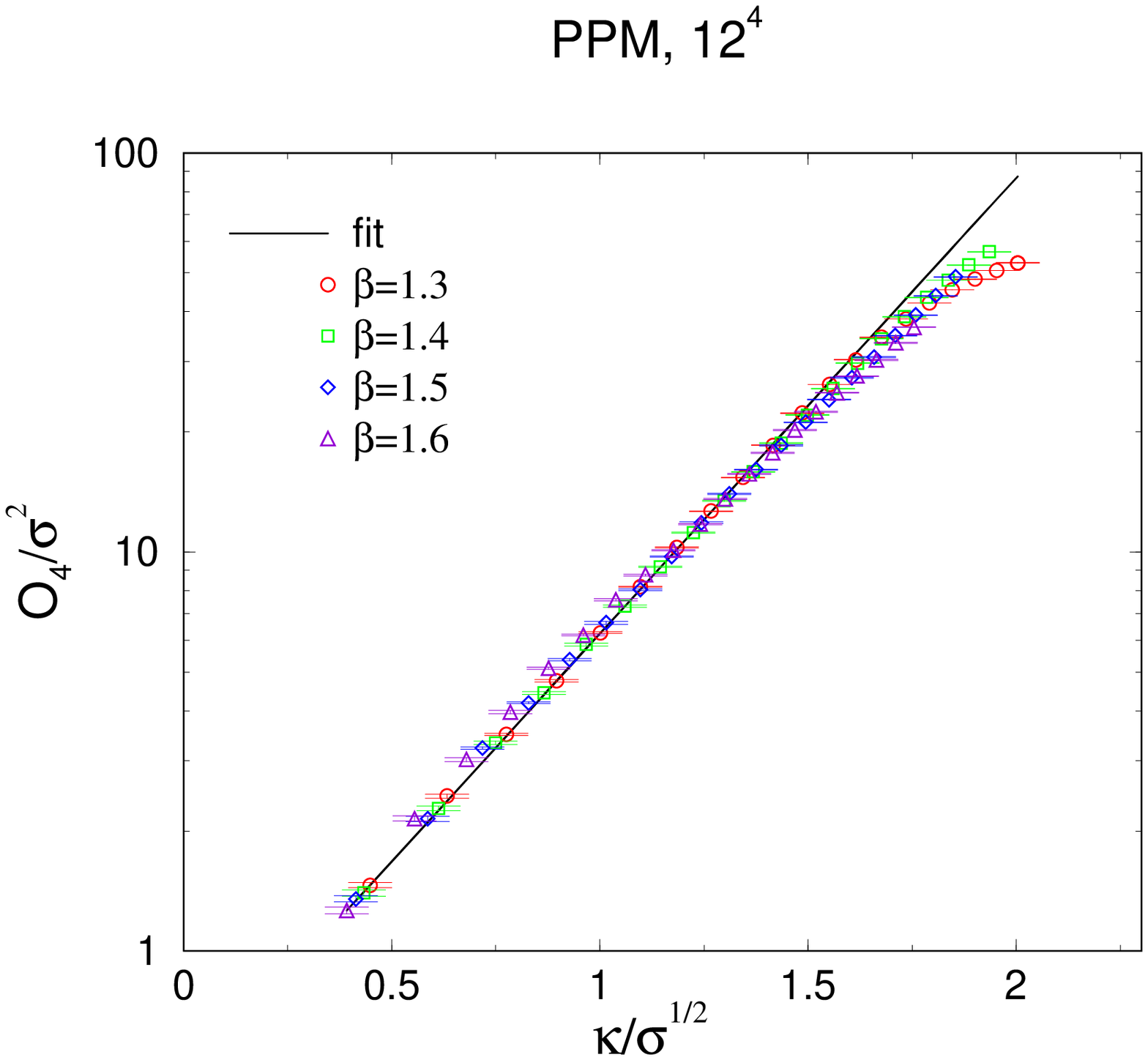}
}
\caption{ The $SU(2)$ action density as function of the cooling 
   scale $\kappa $ in the PPM. } 
\label{fig:8}
\end{figure}
\vskip 0.3cm
Cooling (as well as gauge fixing) is a non--local procedure 
on the link configurations. 
Therefore, it is not excluded that we obtain negative 
plaquettes after $SO(3)$ cooling although we started from a configuration 
with positive plaquettes only\footnote{The same $SO(3)$ action is used 
for the cooling, the positive plaquette constraint is ignored during 
the cooling.}.
Formally repeating the analysis of the previous section for the PPM, 
we present our result for the ``gluon condensate'' 
as function of the cooling scale in figure~\ref{fig:8} (left panel), 
compared with the corresponding result obtained with Wilson action
in a double--logarithmic plot. 
Again the condensate value for each cooling scale 
does not depend on the bare inverse coupling $\beta $ of the PPM
as long as the scaling (\ref{eq:p2}) is applied.
We find in the case of the PPM that the asymptotic value 
(for $\kappa \rightarrow 0$) is non--vanishing, but significantly smaller 
than in the case of the Wilson action. In fact, we find that our data 
for $O_4(\kappa )$ are well reproduced by the fit (see the 
semi--logarithmic plot in figure~\ref{fig:8}, right panel) 
\be 
O_4(\kappa) \; = \; 0.45 \, \sigma ^2 \; \exp \biggl\{  2.63 \, \frac{\kappa }{ 
\sqrt{\sigma } } \biggr\}  \; . 
\label{eq:p5} 
\en 
Extrapolating (\ref{eq:p5}) to $\kappa = 0$, we find that the 
gluon condensate of the PPM is non--vanishing, but one order of 
magnitude smaller than the gluon condensate if the Wilson action 
is used. 

\vskip 0.3cm 
The fact that we have obtained a renormalization group invariant and 
non--vanishing gluon condensate also in the PPM, in which 
the plaquettes are constrained to be positive, 
stirs the hope 
that our above defined gluon condensate
gets physical significance independent of the choice of the lattice 
action. We attribute the fact that we don't find the same value of 
the condensate in the case of Wilson action and in the PPM to the 
deficiency of the latter to match with perturbative  
scaling at very short distance. This is because the gluon condensate 
appears as the first correction to a perturbative calculation.

\section{Conclusions \hfill }

In this paper, we have separated the $SU(2)$ gauge field degrees 
of freedom into thin $Z(2)$ center vortices and $SO(3)$ coset fields. 
Since the $SO(3)$ coset fields are isomorphic to algebra valued fields, 
these degrees of freedom have been identified with the {\it gluonic} 
ones.

\vskip 0.3cm
A new self--restricted cooling algorithm which reduces the $SO(3)$
action of the coset fields facilitates the gradual removal of the 
gluon fields from the lattice configurations while preserving the 
center degrees of freedom. This $SO(3)$ cooling procedure is gauge 
covariant. Hence, the remaining $SU(2)$ Wilson action density reveals 
the gauge invariant $Z(2)$ vortex texture of the $SU(2)$ vacuum. 

\vskip 0.3cm
Extracting the string tension for several values of the gauge invariant 
cooling scale $\kappa/\sqrt{\sigma}$, we have found that the string 
tension vanishes in the limit of unlimited cooling. This shows that 
the $Z(2)$ vortex ensembles remaining after $SO(3)$ cooling 
cannot be identified with the confining $P$--vortices found in the MCG.

\vskip 0.3cm
The operator product expansion (OPE) does not offer 
an unambiguous 
prescription for identifying the condensates. 
We have suggested here an appealing picture: the mass dimension four 
condensate is given by the action density of the effective $Z(2)$ gauge 
model configurations which remain after the $SO(3)$ cooling procedure. 

\vskip 0.3cm
This proposal gets support from the following numerical 
observations: first, the $SU(2)$ action density for a given 
cooling scale properly scales towards the continuum limit; 
second, this action density approaches a renormalization group 
invariant constant in the limit of infinite cooling, 
when the $SU(2)$ field is reduced to its $Z(2)$ (vortex) content. 
This quantity gets immediate importance as the gluon condensate 
figuring in the OPE approach.

\vskip 1cm

{\bf Acknowledgments: }

We thank M.~Engelhardt for critical discussions. 
K.~L.~gratefully acknowledges discussions with G.~Burgio and F.~Di Renzo.

\end{document}